\documentclass[twocolumn,preprintnumbers,secnumarabic,amsmath,amssymb,nofootinbib,floatfix]{revtex4}
\usepackage{varioref,exscale,latexsym,amsmath,amssymb}
\usepackage{graphicx}

\usepackage{graphicx}
\usepackage{bm}
\usepackage{axodraw}

\def\beq{\begin{equation}}

\def\eeq{\end{equation}}

\def\beqa{\begin{eqnarray}}

\def\eeqa{\end{eqnarray}}

\begin{document}

\title{{\bf Comments on the  Minimal Vectorial Standard Model}}

\medskip\
{\author{Mohamed M. Anber}
\email[Email: ]{manber@physics.umass.edu}
 \author{Ufuk Aydemir}
 \email[Email: ]{uaydemir@physics.umass.edu}
\author{ John F. Donoghue}
\email[Email: ]{donoghue@physics.umass.edu}
\author{Preema Pais}
\email[Email: ]{ppais@physics.umass.edu}
\affiliation{Department of Physics,
University of Massachusetts\\
Amherst, MA  01003, USA}

\begin{abstract}
We explore the available parameter space of the minimal vectorial Standard Model. In this theory, the gauge currents are initially vectorial but the Higgs sector produces chiral mass eigenstates, leading to a set of heavy right-handed mirror particles. We describe the phenomenology of the residual parameter space and suggest that the model will be readily tested at the LHC.
\end{abstract}
\maketitle


\section{Introduction}

QCD and QED are vectorial gauge theories - the gauge currents are the same for left-handed and right-handed fermions. However the weak $SU(2)$ currents (and the original $U(1)$ of hypercharge) are chiral with an asymmetry between left and right-handed fields. The $SU(2)$ interaction is maximally asymmetric, coupling to only left-handed fields. However, it is not just that Nature favors left-handed fields because the hypercharge gauge theory carries a complicated mixture of left and right couplings, which conspire to make QED vectorial after symmetry breaking. One has to wonder about the fundamental origin of this unusual asymmetry in handedness \cite{Lee:1956qn}.

Actually there is a variation of the Standard Model that has only vectorial gauge currents. This is {\it not} the $SU(2)_L \times SU(2)_R$ model with separate left and right gauge bosons\cite{Mohapatra}.  Rather the vectorial Standard Model has only the usual set of $SU(2)\times U(1)$ gauge bosons, and all of these couple vectorially to all fermions. It is the Higgs sector which introduces chirality and separates left fields from right. To see how this works, consider an $SU(2)$ doublet field and a pair of singlets.
\begin{equation}
\psi=
\left(\begin{array}{c}
a\\
b
\end{array}\right)\,\,\,,\mbox{}\,\,~~~~~ c, ~~~~f~~.
\end{equation}
The quantum number assignments are the same for both the L,R components, resulting in vectorial gauge currents. However a Higgs doublet will automatically couple the left-handed components of a doublet to right-handed components of a singlet, and the reverse\footnote{There can also be direct mass terms without a Higgs coupling - these are described below.}. If the Higgs couplings are different for these cases, the chiralities will be split after the Higgs picks up a vacuum expectation value. In particular the four vector fields $a,~b,~c,~f$ will be split into four chiral fields, which we can call $u, d, U, D$ which are mass eigenstates with composition
\begin{eqnarray}
u &\sim& (a_L, c_R)  \nonumber \\
U &\sim& (a_R, c_L)  \nonumber \\
d &\sim& (b_L, f_R)  \nonumber \\
D &\sim& (b_R, f_L)\,.
\end{eqnarray}
In this case, $u, d $ will have left-handed weak interactions and $U, D$ will have right-handed ones. With the standard hypercharge assignments, QED will be vectorial for all fields. If the $U,D$ are heavy enough they would not have been uncovered yet.

This vectorial Standard Model was proposed in the lattice QCD literature \cite{Montvay:1987ys, Maalampi:1982ak, Foot:1991bp,Dugan:1991ck, Creutz:1994ny}. In that case it is valuable because of the difficulty of defining chiral gauge theories on a lattice \cite{Nielsen:1981hk}. Having a vector gauge theory avoids this trouble and one then subsequently splits the fermions to produce the usual Standard Model. This in itself is a motivation for this variation of the Standard Model. Our best candidate for a full non-perturbative regularization of a field theory is the lattice, and if there is a fundamental obstacle to chiral gauge theories on a lattice this may turn out to be a more general obstacle with other non-perturbative regularization schemes. Moreover, one can imagine many ways that gauge theories emerge from some discrete underlying theory - these would likely share the lattice obstacle. The Higgs sector is much less understood and it could arguably be more plausible to have the chiral asymmetry in the Higgs sector.

Independent of the favored motivation, the minimal vectorial Standard Model is a construction of some beauty, and it deserves to be tested experimentally. It is clear that we need to make the mirror fields heavy - this is not a problem on its own. However, having heavy chiral fields coupled to the weak currents leads to problems \cite{Dugan:1991ck} with the precision electoweak observables, in particular the $S,T$ parameters\cite{Peskin:1990zt}. Nevertheless, there is still a region of parameter space that satisfies all these constraints - we will explore this below. The allowed region of the parameter space is rather small, so that the model is highly fine-tuned. However when considering further testing of the model this is an advantage - the allowed masses are well constrained. This implies that the model should be quickly confirmed or ruled out at the LHC.

A crucial caveat is that we are exploring this model perturbatively. However, the Yukawa couplings of the mirror fermions are quite large and one has to worry that either higher orders in perturbation theory or non-pertubative effects will modify the standard calculations. At present we have very little idea of the impact of possible non-perturbative effects.

In Section 2 we explore the model more fully using a variant with two Higgs doublets. The electroweak observables are studied in Sec. 3, which identifies the allowed mass ranges. Possible applications to baryogenesis is explored in Sec. 4, with mostly negative conclusions. The use of the mirror leptons as dark matter is describe in Sec. 5, with a favorable result if the possible Majorana masses are adjusted appropriately. Finally we discuss the prospects for testing this model at the LHC in Sec. 6, with mostly favorable conclusions. A summary and two appendices complete the manuscript.

\section{The model}

The model has three $SU(2)$ doublets of quarks
\begin{equation}
\psi_{\alpha}=
\left(\begin{array}{c}
a_\alpha\\
b_\alpha
\end{array}\right)\,\,\,,
\end{equation}
together with singlets $c_\alpha$ and $f_\alpha$. There are also three generations of leptons constructed in a similar fashion. The matter Lagrangian is given by
\begin{equation}
{\cal L}_{m}=i\bar\psi_{\alpha}\displaystyle{\not}D\psi_{\alpha}+i\bar c_{\alpha}\displaystyle{\not}D c_{\alpha}+i\bar f_{\alpha}\displaystyle{\not}D f_{\alpha}\,,
\end{equation}
where
\begin{eqnarray}
\nonumber
D_\mu \psi_{\alpha}&=&\left(\partial_{\mu}+ig_2 \vec \tau \cdot \vec W_\mu/2+ig_1 Y^{\psi} B_{\mu}/2 \right)\psi_{\alpha}\,,\\
\nonumber
D_\mu c_{\alpha}&=&\left(\partial_{\mu}+ig_1 Y^{c} B_{\mu}/2  \right)c_{\alpha}\,,\\
D_\mu f_{\alpha}&=&\left(\partial_{\mu}+ig_1 Y^{f} B_{\mu}/2  \right)f_{\alpha}\,,
\end{eqnarray}
and sums over the Greek indices are implicit. In order to avoid too many new names, we will use the symbols $(a,b,c,f)$ for leptons as well as quarks. The hypercharge assignments are standard and will be stated below once we identify the mass eigenstates.

\subsection*{Higgs sector}

It is possible to construct the model with a single Higgs doublet.  However, the electroweak precession data, as we will see in the next section, requires additional new physics contributions beyond that of a single Higgs doublet. The minimal version of such an extension is to invoke two-Higgs-doublets \cite{Flores:1982pr}
\begin{equation}
{\cal L}_{H}=D_\mu\Phi_1 D^\mu\Phi_1+D_\mu\Phi_2 D^\mu\Phi_2-V(\Phi_1,\Phi_2)\,,
\end{equation}
where
\begin{equation}
D_\mu\Phi_{1,2}=\left(\partial_{\mu}+ig_2 \vec \tau \cdot \vec W_\mu/2+ig_1  B_{\mu}/2 \right)\Phi_{1,2}\,,
\end{equation}
and $V$ is the self-interaction potential
\begin{eqnarray}
\nonumber
V(\Phi_1,\Phi_2)=&-&\mu_1^2\Phi_1^{\dagger}\Phi_1-\mu_2^2\Phi_{2}^{\dagger}\Phi_2+\frac{\lambda_1}{2}\left(\Phi_1^{\dagger}\Phi_1\right)^2\\
\nonumber
&+&\frac{\lambda_2}{2}\left(\Phi_2^{\dagger}\Phi_2\right)^2+h_1\left(\Phi_1^\dagger\Phi_2\right)\left(\Phi_2^\dagger\Phi_1\right)\\
\nonumber
&+&h_2\left(\Phi_1^\dagger\Phi_1\right)\left(\Phi_2^\dagger\Phi_2\right)\\
\label{Vdouble}
&+&h_{3}\left(\left(\Phi_1^\dagger\Phi_2\right)^2+\left(\Phi_2^\dagger\Phi_1\right)^2 \right)\,.
\end{eqnarray}
The parameters $\mu_1\,,\mu_2\,,\lambda_1\,,\lambda_2\,,h_1\,,h_2\,,h_3$ and $h_4$ are taken to be real and are chosen such that the potential is bounded from below.
The Higgs spectrum consists of two charged $H_{\pm}$, CP even $h_0$ and $H_0$, and CP odd $A_0$ fields.

The fermions couple to the two Higgs doublets through Yukawa terms that are invariant under SU(2)$\times$U(1)
\begin{eqnarray}
\nonumber
{\cal L}_{Y}=&-&\Gamma^{1\,c}_{\alpha_\beta}\bar \psi_{\alpha\, L}\tilde \Phi_1 c_{\beta\,R}-\Gamma^{1\,f}_{\alpha_\beta}\bar \psi_{\alpha\, L} \Phi_1 f_{\beta\,R}\\
\label{Yukawa Lagrangian}
&-&\Gamma^{2\,c}_{\alpha_\beta}\bar \psi_{\alpha\, R}\tilde \Phi_2 c_{\beta\,L}-\Gamma^{2\,f}_{\alpha_\beta}\bar \psi_{\alpha\, R} \Phi_2 f_{\beta\,L}+\mbox{h.c.}\,.
\end{eqnarray}
The terms (\ref{Vdouble}) and (\ref{Yukawa Lagrangian}) are invariant under a discrete symmetry of the form
\begin{eqnarray}
\nonumber
&\Phi_1& \longrightarrow \Phi_1\,, \quad \Phi_2\longrightarrow-\Phi_2\\
\nonumber
&\psi_R&\longrightarrow \psi_R\,, \quad \psi_L\longrightarrow-\psi_L\\
\nonumber
&c_R&\longrightarrow-c_R\,, \quad c_L\longrightarrow -c_L\\
\label{the discrete symmetry}
&f_R&\longrightarrow-f_R\,, \quad f_L\longrightarrow -f_L\,.
\end{eqnarray}
which prevents the mixing between $\Phi_1$ and $\Phi_2$, and in the same time suppresses flavor changing neutral current (FCNC). In fact, there are 14 possible discrete symmetries which are possible for this model, and which could impact the properties of the fermion mass terms. These are described in Appendix A.

After spontaneously breaking the SU(2) symmetry, the Higgs fields acquire VEVs
\begin{eqnarray}
\left\langle \Phi_1 \right\rangle=
\frac{1}{2}\left(\begin{array}{c}
0\\
v_1
\end{array}\right)\,, \quad \left\langle \Phi_2 \right\rangle=
\frac{1}{2}\left(\begin{array}{c}
0\\
v_2 e^{i\theta_2}
\end{array}\right)\,,
\end{eqnarray}
which preserve the $U(1)$ gauge symmetry.

\subsection*{Mass spectrum from the Higgs sector}

First consider the spectrum without direct mass terms. This can be arranged by a discrete symmetry - see Appendix A. In this case we emerge with the usual left handed particles and a mirror set with right handed couplings.

The Yukawa-coupling matrices may be diagonalized by bi-unitary transformations
\begin{eqnarray}
\nonumber
a_R&=&V_{R}U_R\,, \quad a_L=V_Lu_L\\
\nonumber
b_R&=&X_{R}D_R\,, \quad b_L=X_Ld_L\\
\nonumber
c_R&=&S_{R}u_R\,, \quad c_L=S_LU_L\\
\label{BiUnitary trans}
f_R&=&W_{R}d_R\,, \quad f_L=W_LD_L\,,
\end{eqnarray}
where $u_{L,R}$ ($U_{L,R}$) and $d_{L,R}$ ($D_{L,R}$) denote the $3\times 1$ column matrices with the chiral components of the fermion mass eigenstates of the physical Standard Model (mirror) fermions. The $3\times 3$ unitary matrices $V$, $X$, $S$, $W$ are chosen to bi-diagonalize the physical mass matrices $M_{d}$, $M_{u}$, $M_{D}$ and $M_{U}$
\begin{equation}
\begin{array}{cc}
M_{u}=v_1V^{\dagger}_{L}\Gamma^{1\,c}S_{R}/2& M_{d}=v_1 X^{\dagger}_{L}\Gamma^{1\,f}W_{R}/2 \\
M_{U}=v_2e^{-i\theta_2} V^{\dagger}_{R}\Gamma^{2\,c}S_{L}/2 & M_{D}=v_2e^{-i\theta_2} X^{\dagger}_{R}\Gamma^{2\,f}W_{L}/2\,.
\end{array}
\end{equation}
If $v_2>>v_1$, then our choice for the Yukawa Lagrangian (\ref{Yukawa Lagrangian}) can explain why the mirror masses are naturally heavier than the SM ones.

The charges of these particles are given by $Q_{el}=Y_w/2+T_{w3}$, where $T_{w3}$ denotes the third component of the weak isospin, and $Y_w$ is the corresponding hypercharge (see table (\ref{assignments}) for the weak isospin and hypercharge assignments.)
\begin{table}
\centerline{
\begin{tabular}{|cccc|cccc|}
	\hline
\mbox{quarks} & $T_w$ & $T_{w3}$  & $Y_w$ &  leptons & $T_w$ & $T_{w3}$  & $Y_w$\\
	\hline
	\hline
	$a_q$,\,$u_{q\,L}$,\,$U_{q\,L}$ & $1/2$ & $1/2$ & $1/3$ &  $a_l$,\,$u_{l\,L}$,\,$U_{l\,L}$ & $1/2$ & $1/2$ & $-1$\\
	$b_q$,\,$d_{q\,L}$,\,$D_{q\,L}$ & $1/2$ & $-1/2$ & $1/3$ & $b_l$,\,$d_{l\,L}$,\,$D_{l\,L}$ & $1/2$ & $-1/2$ & $-1$\\
	$c_q$,\,$u_{q\,R}$,\,$U_{q\,R}$ & $0$ & $0$ & $4/3$ & $c_l$,\,$u_{l\,R}$\,,$U_{l\,R}$ & $0$ & $0$ & $0$ \\
	$f_q$,\,$d_{q\,R}$,\,$D_{q\,R}$ & $0$ & $0$ & $-2/3$ &	 $f_l$,\,$d_{l\,R}$,\,$D_{l\,R}$ & $0$ & $0$ & $-2$ \\		
	\hline		
\end{tabular}
}
\caption{$SU(2)\times U(1)$ assignments for the SM and mirror particles.}
\label{assignments}
\end{table}

The charged current interaction  is
\begin{eqnarray}
\nonumber
{\cal L}_{W}=&-&\frac{g_2}{2\sqrt{2}}W^{-\,\mu}\left[\bar d \gamma_{\mu}V^{CKM}(1+\gamma_5)u\right.\\
&+&\left.\bar D\gamma_{\mu}V^{M\,CKM}(1-\gamma_5)U \right]+\mbox{h.c.}\,,
\end{eqnarray}
where
\begin{equation}
\begin{array}{cc}
V^{CKM}=X_L^\dagger V_L\,, & V^{M\,CKM}=X_R^\dagger V_R\\
\end{array}
\end{equation}
are Cabibbo-Kobayashi-Maskawa matrices for the SM and mirror fermions, respectively. The electromagnetic and weak neutral currents are standard, with the obvious change that the neutral current of the mirrors involves their right handed components rather than left.

\subsection*{Mixing via direct mass terms}

In addition to the Yukawa couplings in (\ref{Yukawa Lagrangian}), we can also introduce bare mass terms that are invariant under the SU(2) symmetry. While some discrete symmetries forbid these terms, there are others that allow specific combinations.  For example, the mass terms that are allowed by the discrete symmetries which are called S1, S2, S4 or S5 given in Table (\ref{symmetries}) in Appendix A takes the form
\begin{equation}\label{bare Lagrangian cf}
{\cal L}_{bare,cf}=-m_{c\alpha} \bar c_\alpha c_\alpha-m_{f\alpha} \bar f_\alpha f_\alpha\, .
\end{equation}
However a different mass structure is possible under different discrete symmetries. Those labeled S3 or S6 in Appendix A allow only
\begin{equation}\label{bare Lagrangian psi}
{\cal L}_{bare,\psi}=-m_{\psi,\alpha} \bar \psi_\alpha \psi_\alpha\,.
\end{equation}
In principle the scale of these mass terms could take on any value. However, we will treat these masses as smaller than the Higgs generated masses - this is required if the observed quarks have dominantly left-handed weak interactions. Since the $SU(2)$ interaction is still weakly coupled at this scale, one might expect that the masses could scale as $ v e^{-8\pi^2/g^2_2}\sim 10^{-69}$~eV, although we will not make any specific assumption about that mass scale in this work.

Using the bi-unitary transformations in Eq. (\ref{BiUnitary trans}) can provide a further diagonalization of the masses in order to obtain
\begin{eqnarray}
\nonumber
{\cal L}_{\mbox{\scriptsize bare},cf}=-\frac{1}{2}\bar U K_{1}\left(1 -\gamma_5 \right)u-\frac{1}{2}\bar D K_{2}\left(1 +\gamma_5 \right)d+\mbox{h.c.}\,,\\
\nonumber
{\cal L}_{\mbox{\scriptsize bare},\psi}=-\frac{1}{2}\bar U K_{3}\left(1 +\gamma_5 \right)u-\frac{1}{2}\bar D K_{4}\left(1 -\gamma_5 \right)d+\mbox{h.c.}\,,\\
\label{mixing Lagrangian}
\end{eqnarray}
where
\begin{eqnarray}
\nonumber
K_{1}&=& S_L^\dagger m_c S_R\,, \quad K_{2}= W_R^\dagger m_f W_L\\
K_{3}&=& V_R^\dagger m_\psi V_L\,, \quad K_{4}= X_L^\dagger m_\psi X_R\,.
\end{eqnarray}

The effect of the mixing Lagrangian above (\ref{mixing Lagrangian}) is that it leads to mixing of the chiral eigenstates of the generic form
\begin{equation}\label{quark mixing}
u_{\mbox{\scriptsize phys}}= u + \frac{K_1}{m_U -m_u}\, U
\end{equation}
to first order. If such mixing exists, this will allow the decays of mirror matter into normal matter.
If the scale of the mass terms is $10^{-69}$~eV, the mixing between normal and mirror fields will be extremely tiny, of the order $\theta_{mix} \sim 10^{-80}$. This would make the mirrors essentially stable.

\subsection*{Majorana mass terms}

In addition to the Dirac mass terms, the neutrinos can also have Majorana mass terms. The Majorana mass terms that are invariant under the $SU(2)\times U(1)$ symmetry may be written
\begin{equation}\label{majorana L}
{\cal L}_{\mbox{\scriptsize Majorana}}=\frac{1}{2}c_R^T C^{-1} M_R^\nu c_R+\frac{1}{2}c_L^T C^{-1} M_L^N c_L+\mbox{h.c.}\,,
\end{equation}
where $C$ is the charge conjugation operator, and $M_R^\nu$ and $M_L^N$ are $3\times 3$ matrices. Using the bi-unitary transformation (\ref{BiUnitary trans}), we find
\begin{equation}
{\cal L}_{\mbox{\scriptsize Majorana}}=\frac{1}{2}\nu_R^T C^{-1}{\cal M}_R^\nu\nu_R+\frac{1}{2}N_L^T C^{-1}{\cal M}_L^N N_L+\mbox{h.c.}\,,
\end{equation}
where
\begin{equation}
{\cal M}_L^N=S_L^TM_L^N S_L\,, \quad {\cal M}_R^\nu=S_R^TM_R^\nu S_R\,.
\end{equation}

At this point, one can combine direct, Dirac and Majorana mass terms
\begin{eqnarray}
\nonumber
& &{\cal L}_{\mbox{\scriptsize mass}}=-\frac{1}{2} \\
& &
\left[\begin{array}{cccc}\bar{\nu_L^c} & \bar{\nu_R} & \bar{N_L^c} & \bar{N_R} \end{array}\right]\left[\begin{array}{cccc}0 & M_D^{\nu} &0 & K_3\\ M_D^{\nu\,T} & {\cal M}_R^\nu &K_1 &0\\
0 &K_1 & {\cal M}_L^N & M_D^{N}\\
K_3& 0 & M_D^{N\,T} & 0\end{array} \right]\left[\begin{array}{c}\nu_L \\ \nu_R^c \\N_L \\ N_R^c \end{array}\right] \nonumber \\
\end{eqnarray}
where $M_D$ denotes the Dirac mass matrix. We obtain the usual seesaw mechanism  we consider for the left-handed neutrino in the limit ${\cal M}_R^\nu >>M_D$, which has the potential to explain the smallness of the SM neutrino mass.  For the mirror $N$, we would favor ${\cal M}_L^N <M_D$ so that the neutral mirror is not {\em too} light.

 Note that the gauge singlet neutral particles will also mix with heavier singlet particles. In the usual version of the Standard Model, one invokes extra singlets outside of the SM in order to implement the seesaw mechanism. These extra singlets are often given large masses $\sim 10^{10}$~GeV which produces neutrino masses of the right size if the Dirac masses are also comparable to those of the charged leptons. This option is available in the vectorial SM also. However, one need not go outside of the vectorial SM fields to implement the seesaw mechanism. On the other hand, there does not appear to be a natural explanation of the sizes of the Majorana masses needed in this context.

\section{Fitting the electroweak precision data with mirror particles and two Higgs doublets}

The oblique corrections, parameterized in terms of the $S$, $T$ and $U$ parameters, are extracted from the electroweak precision data, and used to constraint new physics beyond the Standard Model \cite{Peskin:1990zt}. The one-loop fermionic and two-Higgs doublet contributions to these parameters are given in \cite{Haber:1993wf, He:2001tp}. The fermionic contributions to the $S$ parameter are generically quite large if the fermions all have equal masses. This will require a specific hierarchy in the masses in order to be compatible with the data. The one-loop fermionic contributions to the $S, T, U$ parameters are given by
\begin{eqnarray}
\nonumber
S_f&=&\frac{N_c}{6\pi}\left\{2(2Y+3)x_1+2(-2Y+3)x_2-Y\ln(x_1/x_2)\right.\\
\nonumber
&+&\left[(3/2+Y)x_1+Y/2 \right]G(x_1)\\
 &+&\left.\left[(3/2-Y)x_2-Y/2 \right]G(x_2) \right\}\,,
\end{eqnarray}
\begin{equation}
T_f=\frac{N_c}{8\pi \sin^2(\theta_w)\cos^2(\theta_w)}F(x_1,x_2)\,,
\end{equation}
and
\begin{eqnarray}
\nonumber
U_f=&-&\frac{N_c}{2\pi}\left\{\frac{(x_1+x_2)}{2}-\frac{(x_1-x_2)^2}{3}\right.\\
\nonumber
&+&\left[\frac{(x_1-x_2)^3}{6}-\frac{(x_1^2+x_2^2)}{2(x_1-x_2)}  \right]\ln(x_1/x_2)\\
\nonumber
&+&\frac{x_1-1}{6}f(x_1,x_1)+\frac{x_2-1}{6}f(x_2,x_2)\\
\nonumber
&+&\left. \left[\frac{1}{3}-\frac{x_1+x_2}{6}-\frac{(x_1-x_2)^2}{6}  \right]f(x_1,x_2)\right\}\,,\\
\end{eqnarray}
where $x_i=(M_i/M_Z)^2$ with $i=1,2$ and the color factor $N_c=3(1)$ for quarks (leptons). We assign $(M_1,M_2)\longleftrightarrow (M_N,M_E)$ for leptons and  $(M_1,M_2)\longleftrightarrow (M_U,M_D)$ for quarks, and we have $Y=1/3\,(-1)$ for quarks (leptons). The functions  $F(x_1,x_2)$, $G(x)$ and $f(x_1,x_2)$ are given by
\begin{eqnarray}
\nonumber
F(x_1,x_2)&=&\frac{x_1+x_2}{2}-\frac{x_1x_2}{x_1-x_2}\ln(x_1/x_2)\,,\\
G(x)&=&-4\sqrt{4x-1}\arctan\frac{1}{\sqrt{4x-1}}\,,
\end{eqnarray}
and $f(x_1,x_2)=$
\begin{eqnarray}
\left\{
\begin{array}{cc}
-2\sqrt{\Delta}\left[ \arctan\frac{x_1-x_2+1}{\sqrt{\Delta}}-\arctan\frac{x_1-x_2-1}{\sqrt{\Delta}} \right]  & \,,\Delta>0\\
0 &\,, \Delta=0\\
\nonumber
\sqrt{-\Delta}\ln\frac{x_1+x_2-1+\sqrt{-\Delta}}{x_1+x_2-1-\sqrt{-\Delta}} &\,,\Delta<0
\end{array}\right.\\
\end{eqnarray}
where $\Delta=2(x_1+x_2)-(x_1-x_2)^2-1$.
In the limit $M^2_{1,2}>>M_Z^2$ the $S$ parameter approximates to
\begin{eqnarray}
\nonumber
S_f &\approx&\frac{N_c}{6\pi}\left[1-Y\ln\left(\frac{M_1}{M_2}\right)^2+\frac{1+4Y}{20}\left(\frac{M_Z}{M_1} \right)^2\right.\\
&+&\left.\frac{1-4Y}{20}\left(\frac{M_Z}{M_2} \right)^2 \right]\,,
\end{eqnarray}
which in turn reduces to $N_c/6\pi$ for small mass splitting. The most important feature of this model is that the equal-mass limit produces too large a contribution to the S parameter. This requires significant splitting in the fermion masses, as well as splitting in the Higgs sector.

The contribution from the two Higgs doublet is more complicated and given in \cite{He:2001tp}. An example of these to the $T$ parameter in the limit $m^2_{\mbox{\scriptsize Higgs}}>>M_Z^2$
\begin{eqnarray}
\nonumber
T_H&\approx&\frac{1}{16\pi \sin^2(\theta_w)m_W^2}\left\{ \cos^2(\beta-\alpha)\left[F(m^2_{H_{\pm}},m_h^2)\right.\right.\\
\nonumber
&+&\left.F(m^2_{H_{\pm}},m_A^2)-F(m^2_A,m_h^2) \right]\\
\nonumber
&+&\sin^2(\beta-\alpha)\left[F(m^2_{H_{\pm}},m_H^2)+F(m^2_{H_{\pm}},m_A^2)\right.\\
&-&\left.\left.F(m^2_A,m_H^2)\right]
\right\}\,,
\end{eqnarray}
where $\tan\beta=|\left\langle \Phi_2\right\rangle/\left\langle \Phi_1\right\rangle|$, and $\alpha$ is the rotation angle in the the $h_0-H_0$ space.
Further, choosing $\alpha\approx\beta$, and considering the Higgs masses in the range $m_h\approx m_H << m_{H\pm} << m_A$ we obtain
\begin{equation}
T_H\approx\frac{m_{H_\pm}^2}{16\pi \sin^2(\theta_w)m_W^2}\left[1-\ln\left(\frac{m_A^2}{m_{H_\pm}^2}\right) \right]\,.
\end{equation}
%

\begin{figure}
\leftline{
\includegraphics[width=.5 \textwidth]{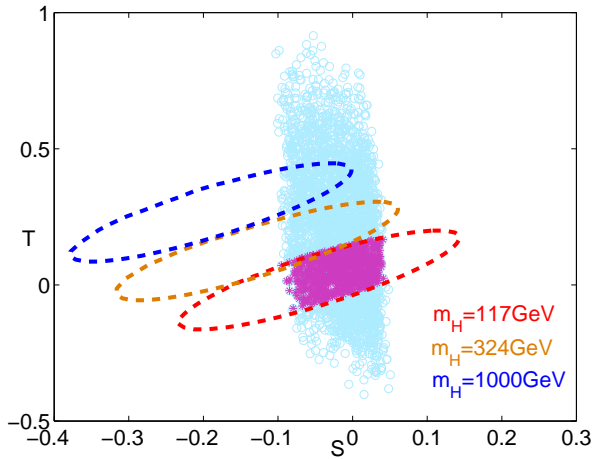}
}
\caption{ The contributions to the electroweak parameters is shown for a sample of $3000$ randomly distributed models as described in the text. The violet region highlights those that satisfy the constraints with a light Higgs. The $U=0$, $90\%$ C.L. contours are shown assuming SM Higgs masses $M_H=117$, $340$ and  $1000$ GeV.}
\label{flowchart}
\end{figure}

\begin{figure*}[ht]
\leftline{
\includegraphics[width=.5\textwidth]{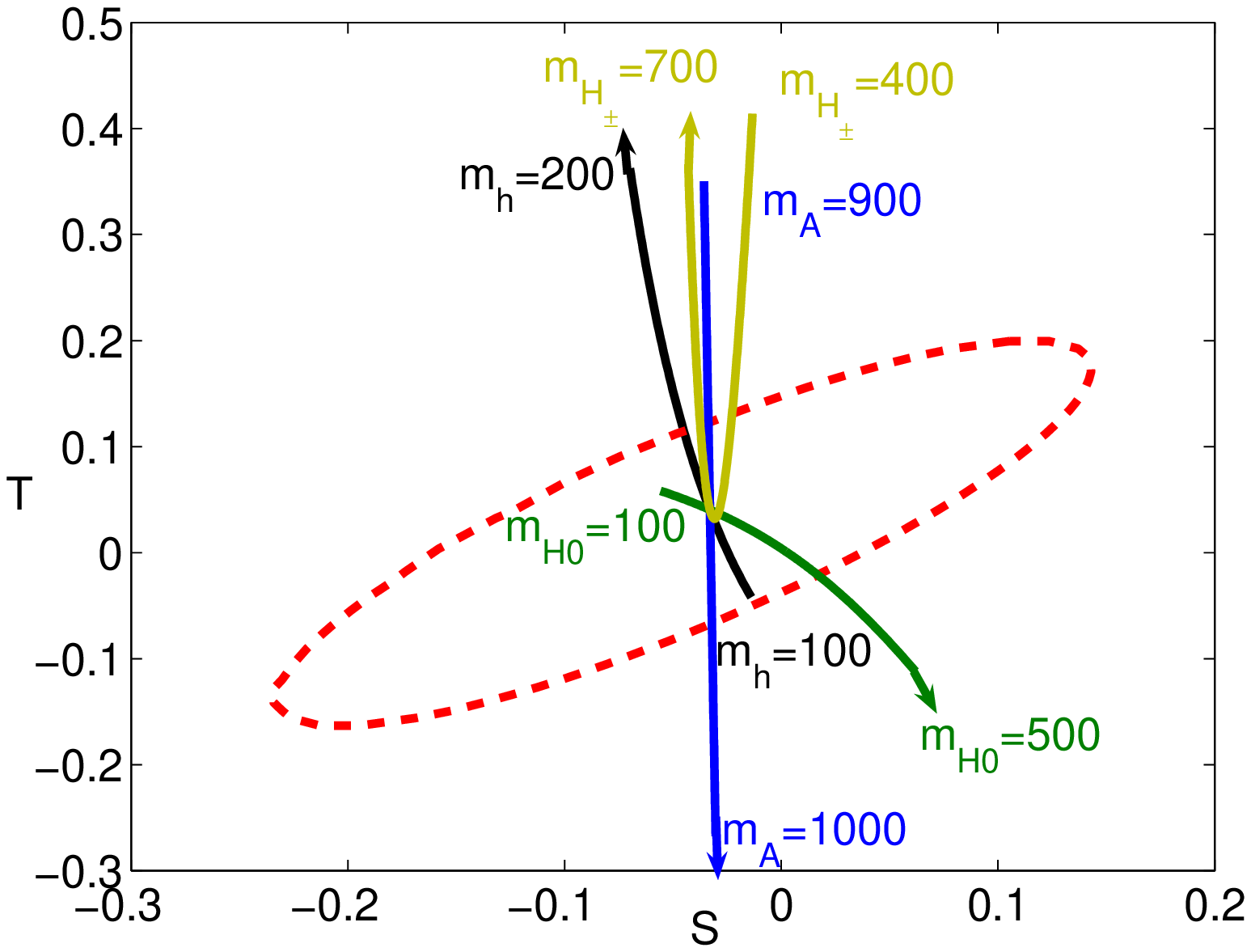}
\includegraphics[width=.5\textwidth]{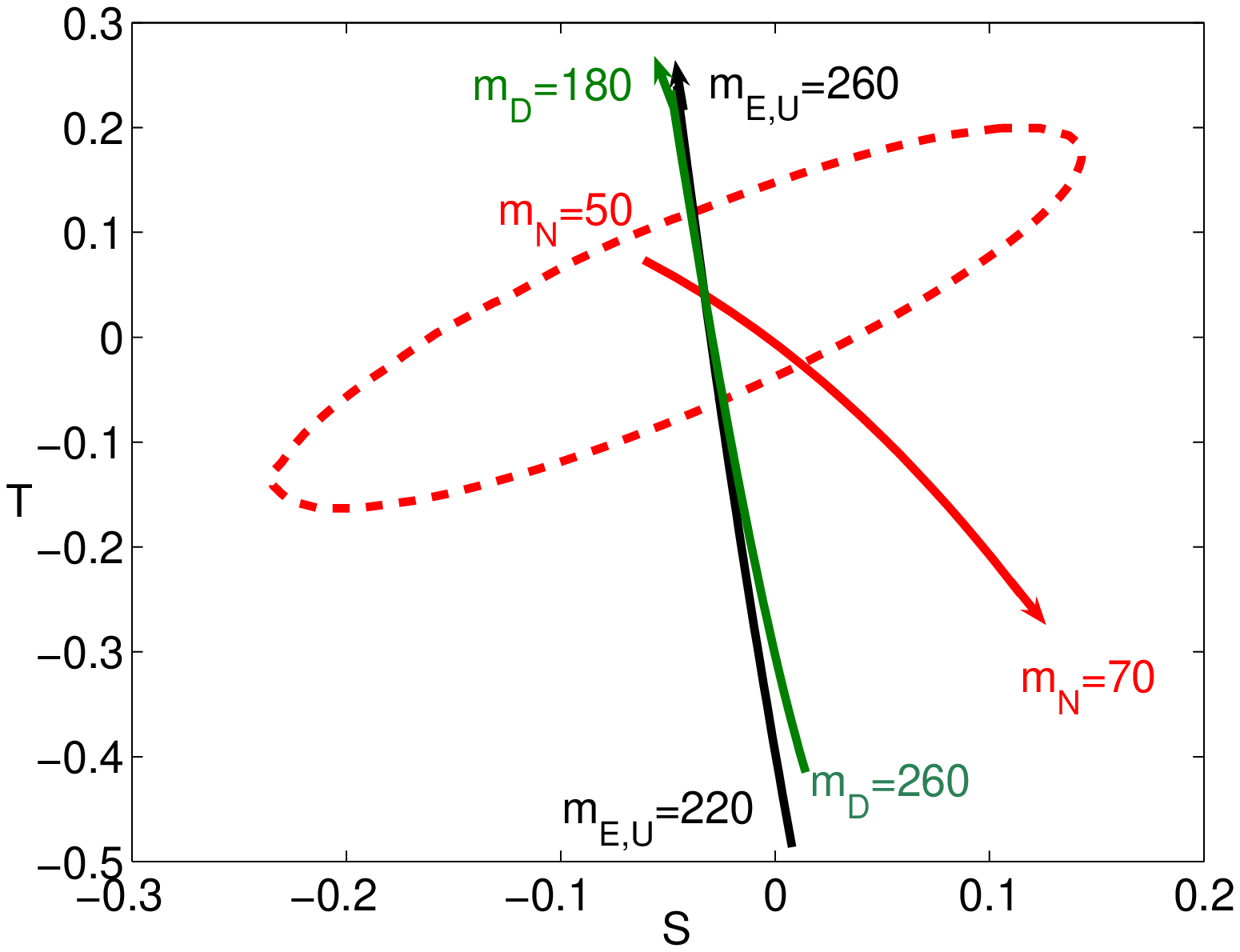}
}
\caption{The effect the variation of the Higgs (on the left) and mirror fermion (on the right) masses  has on the $S$, $T$, $U$ parameters. All curves intersect at $(m_h,m_A,m_H,m_{H\pm})=(125,900,130,580)$ GeV and  $(m_N,m_E,m_U,m_D)=(53,250,250,190)$ GeV. }
\label{dependence}
\end{figure*}

The lepton contribution to $S$ grows with an increasing
$m_{N}$  and with a decreasing $m_{E}$, while the quark contribution behaves in the
opposite way, with smaller values of $S$ correlated with lighter $U$ quark masses. The parameters $T$ and $U$ measure the weak-isospin violation in the SU(2) doublet and are non-vanishing  for $m_N\neq m_E$ or $m_U \neq m_D$. The larger the split between the up and down components the larger their contributions to the $T$ and $U$ parameters. For the heavy right-handed neutrino $N$, there is a firm lower bound on the mass of $45$~GeV from LEP constraints forbidding $Z\to N \bar{N}$. If the $N$ decays through mixing (to be discussed later) , LEPII constraints raise this lower bound to $100$~GeV. However, this can be avoided if the $N$ is stable or almost so. However, one cannot satisfy the electroweak constraints simply by splitting the masses of the heavy mirror particles. 
Canceling the $S$, $T$ and $U$ contribution from three generations of mirror quarks and leptons requires additional new physics contributions, beyond that of a single Higgs doublet, to the oblique parameters. This can be achieved, for example, invoking two-Higgs doublets.

Negative values for the $S$ parameter can be achieved by splitting the up and down values of the mirror leptons and quark masses. This in turn will contribute large positive values for  $T$, and relatively small positive values for $U$. A negative contribution to $T$ as well as negligible effects on $S$ and $U$  in the Higgs sector can always be achieved by choosing the Higgs spectrum in the range $m_h\approx m_H<<m_{H\pm}<<m_A$. If we take $N$ to be the lightest mirror particle (LMP), then we find that starting with the initial values
\begin{equation}\label{initial spectrum}
\left[
\begin{array}{c}
m_N\\
m_E=4m_N
\end{array}
\right]\,,
\left[
\begin{array}{c}
m_U=4m_N\\
m_D=3.2m_N
\end{array}
\right]
\end{equation}
one can use a simple algorithm to fit the $S$, $T$, $U$ parameters to the experimental  data. For simplicity, we have assumed the second and third families to have the same mass spectrum as the first family.  Fig. \ref{flowchart} displays the allowed points in the $S - T$ plane for comparison with the $U=0$, $90\%$ C.L. experimental bounds. Although the $U>0$ contours are not shown, variations in $U$ mainly shift the $S - T$ contour without affecting its shape and direction, and a larger positive $U$ tends to diminish the allowed regions of positive $S$, $T$. The calculations are based on a  sample of $3000$ models randomly distributed  between $m_{\mbox{initial}}$ and $m_{\mbox{final}}=m_{\mbox{initial}}+10$ GeV. The initial values of the fermion masses are taken to be $(m_N,m_E,m_U,m_D)=(50,250,250,190)$ GeV, while the Higgs masses are  $(m_h,m_A,m_H,m_{H\pm})=(125,900,130,580)$ GeV. In all these models we find $0.3<U<0.4$ which restricts the allowed models to those on the left edge of the violet region shown in figure. We have also used the special values $\beta=\alpha=\pi/4$. However, similar results are obtained for the case of $\left\langle \Phi_2\right\rangle\neq\left\langle \Phi_1\right\rangle$.

\begin{figure*}[ht]
\leftline{
\includegraphics[width=.5\textwidth]{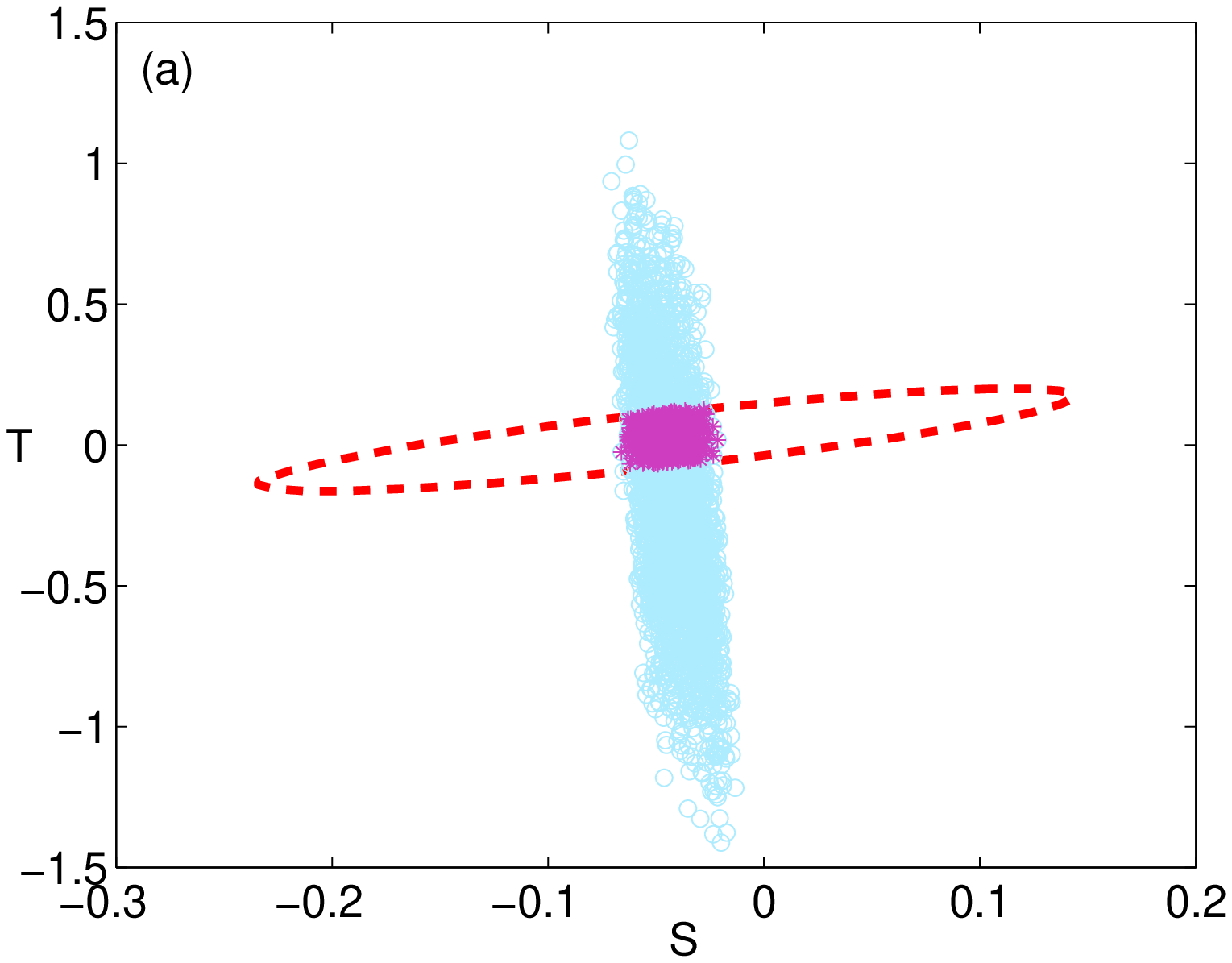}
\includegraphics[width=.5\textwidth]{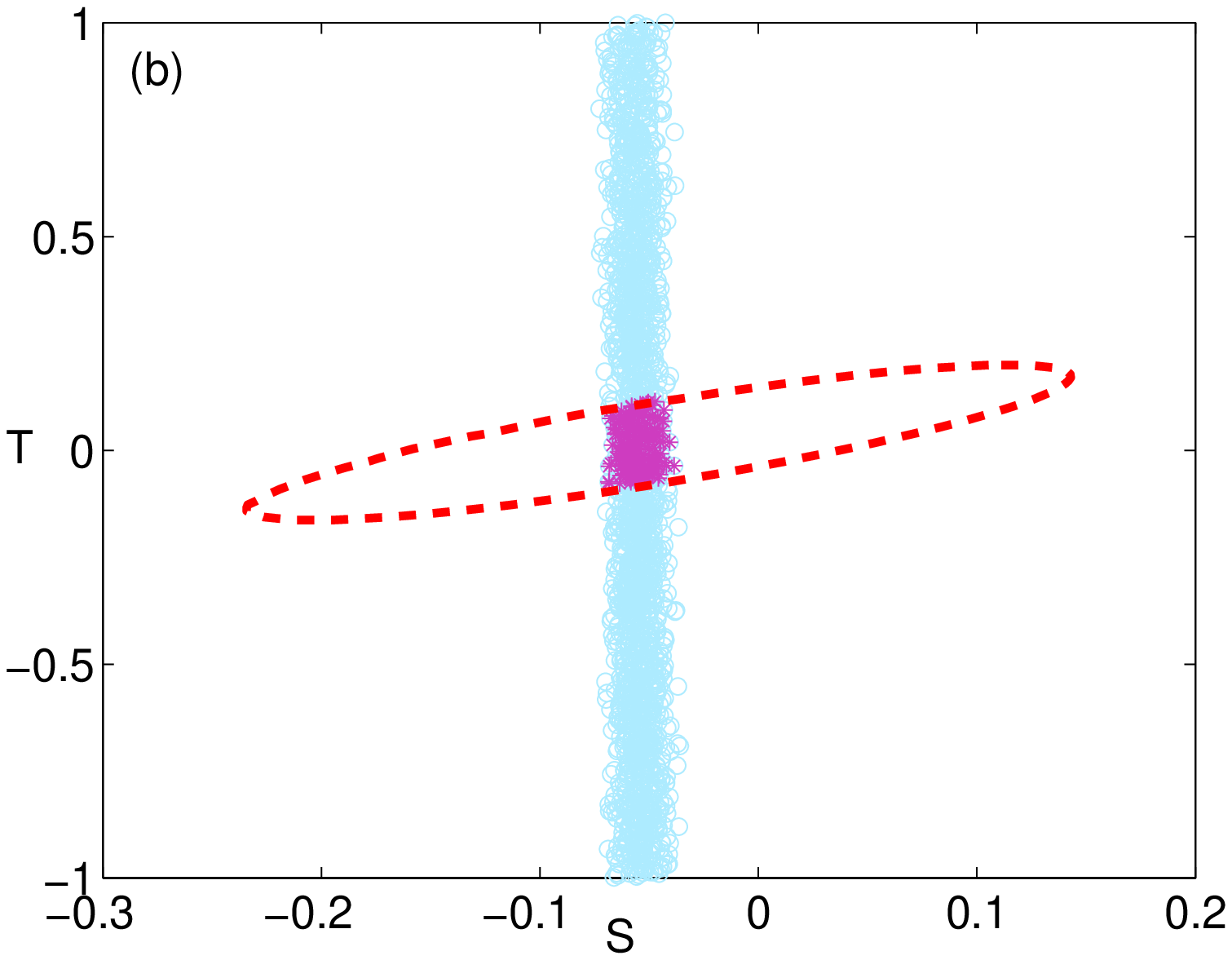}
}
\caption{ The allowed region  (in violet) in a sample of $3000$ randomly distributed models (in light blue) between $m_{\mbox{\scriptsize initial}}$ and $m_{\mbox{\scriptsize final}}=m_{\mbox{\scriptsize initial}} +10$ GeV. The initial values of the fermion  and Higgs masses are taken to be: (a) $(m_N,m_E,m_U,m_D)=(100,580,580,320)$, $(m_h,m_A,m_H,m_{H\pm})=(120,4520,120,800)$ GeV , and (b) $(m_N,m_E,m_U,m_D)=(260,1500,1500,830)$ and $(m_h,m_A,m_H,m_{H\pm})=(115,9035,120,2400)$ GeV.  In both cases we find $U\approx 0.4$.}
\label{highermass}
\end{figure*}

Fig. \ref{dependence} shows the effect the variation of the Higgs and mirror fermion masses has on the $S$, $T$, $U$ parameters. While varying the mass of the Higgs almost does not alter the value of $S$, we find that it has a significant effect on $T$ which is more evident in the case of $m_A$ and $m_{H\pm}$. This is in accordance with the two-independent module algorithm discussed above. Using this algorithm, it is shown in Fig. \ref{highermass} that one can accommodate a LMP as large as $260$ GeV within the experimental constraints of the electroweak precession data.

The existing parameter space of the model then always has the lightest mirror particle being the neutral $N$, with a range of possibilities from $50~{\rm GeV} <m_N <260~{GeV}$. The lighter end of the range is easier to accommodate in the electroweak parameters. The charged mirror leptons $E$ are always much heavier. In the quark sector, the mirror $D$ is always the lightest. In the Higgs sector there is always a large splitting among the physical Higgs bosons, with the lightest being neutral, although there is a significant variation in the ordering and masses of the heavier Higgs.

\section{Electroweak baryogenesis}

\begin{figure*}[ht]
\leftline{
\includegraphics[width=.5\textwidth]{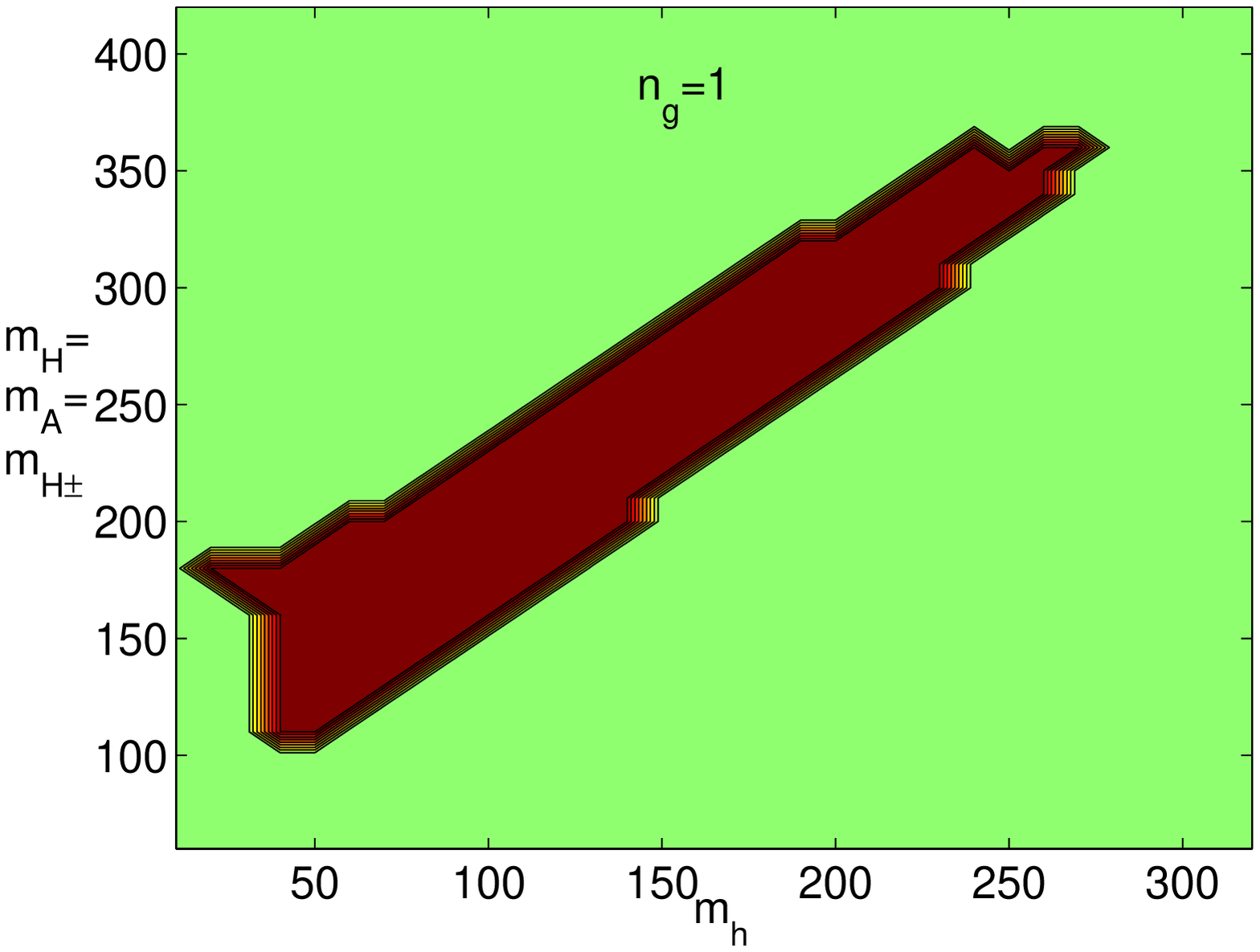}
\includegraphics[width=.5\textwidth]{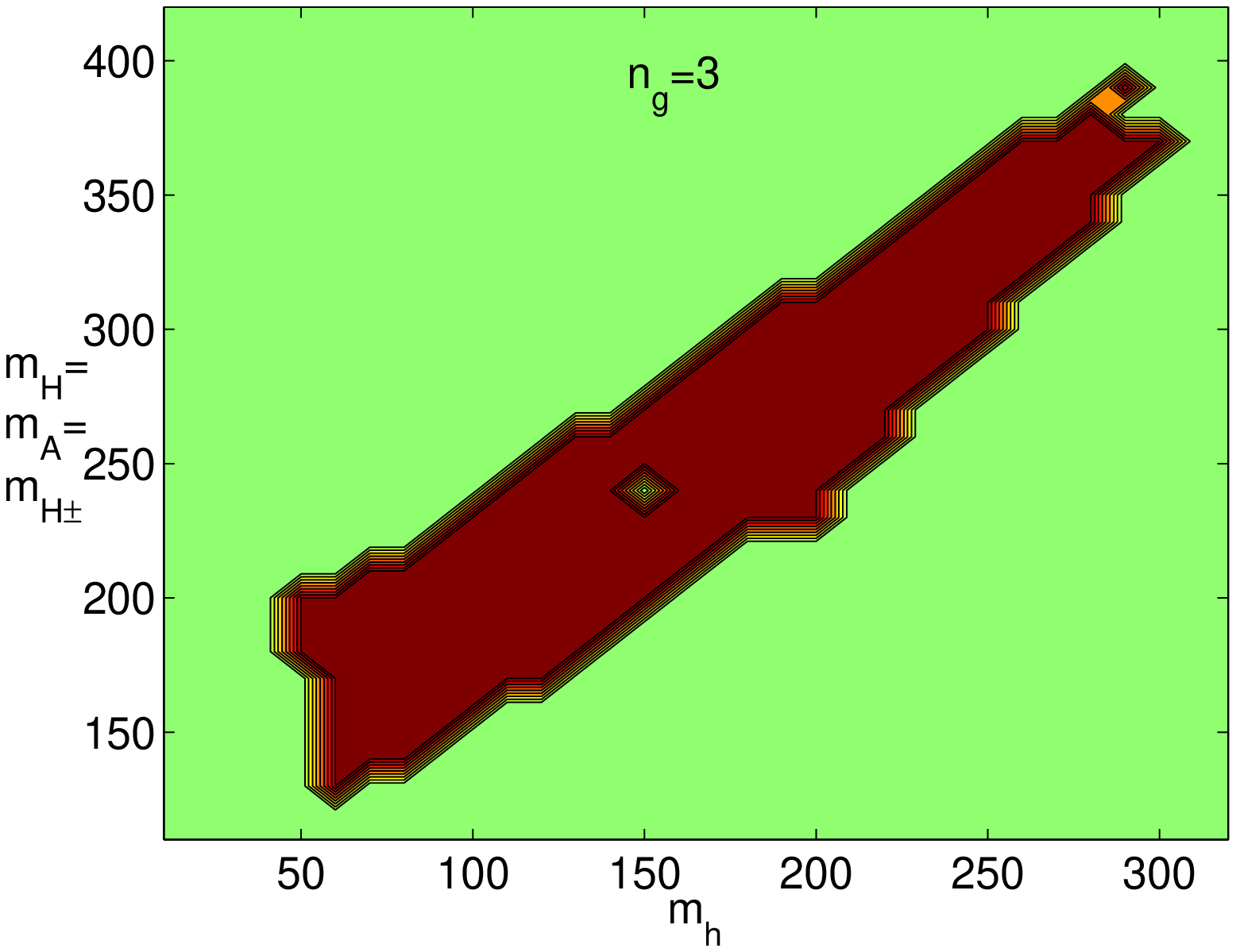}
}
\caption{contours of constant $v_{c}(T_c)/T_c$. Dark regions represent the parameter space with $v_{c}(T_c)/T_c>1$ and $M/T_{c}<1.6$, for which the high temperature expansion is trusted. Units are GeV for masses, and we take one (on the left) and three (on the right) families of mirror particles. We use $m_t=175$ GeV for the top quark, and $m_{N}=m_{E}=m_{U}=m_{D}=100$ GeV for mirror fermions.}
\label{contor}
\end{figure*}

\begin{figure}[ht]
\centerline{
\includegraphics[width=.5\textwidth]{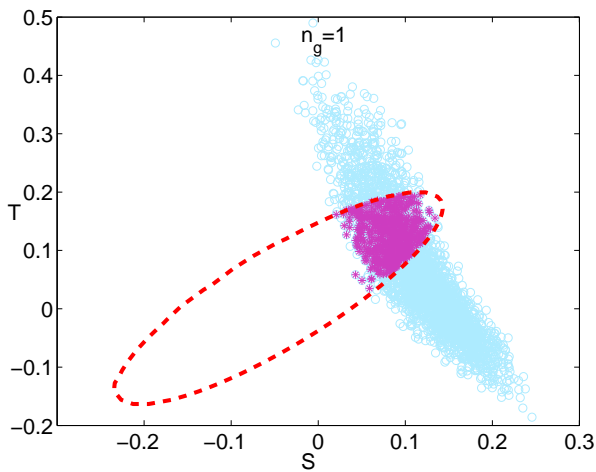}
}
\caption{The allowed region  (in violet) for a sample of $3000$ randomly distributed models (in light blue) using only one generation of mirror particles. All these models can lead to baryogenesis. However, only the points in red represent those  that respect the electroweak precession data (we find that the models in violet have $0<U<0.1$). The $U=0$, $90\%$ C.L. contour is shown assuming SM Higgs masses $M_H=117$ GeV.}
\label{STPLANE}
\end{figure}

Two Higgs doublet models, treated generally, can explain baryogenesis at the electroweak scale \cite{Cline:2006ts, Bochkarev:1990fx, Cline:1995dg}. Here we study whether the mass constraints of the previous section are compatible with baryogenesis. Our answer is negative.

We start by considering the Higgs self potential given in (\ref{Vdouble}). We take $\mu_1^2=\mu_2^2$ and $\lambda_1=\lambda_2$ to simplify the analysis. With this choice, the symmetry will break along the direction $\langle\Phi_1\rangle=\langle\Phi_2\rangle=(0,v)^T/2$, where $v=\sqrt{\mu_1^2/\lambda}=246$ GeV is the tree level VEV, and $\lambda=\left(\lambda_1+h_1+h_2+2h_3 \right)/4$. However, we find that this choice does not change the conclusion we draw below. In appendix A, we work out the details for the electroweak baryogenesis in our vectorial version of the Standard Model.

For electroweak baryogenesis to work, it is necessary that the baryon-violating interactions induced by electroweak sphalerons be sufficiently slow immediately after the phase transition to avoid the destruction of the baryons that have just been created. This condition is fulfilled if the ratio $v_{c}(T_c)/T_c$, the Higgs VEV to the critical temperature at the time of transition, is greater than 1. This ratio is a measure of the strength of the phase transition.

In Fig. \ref{contor} we plot the contours of the ratio $v_{c}(T_c)/T_c$ in the plane of $m_{A}=m_{H}=m_{H_\pm}$ versus $m_{h}$ for one and three families of mirror particles. We apply a cut off for the parameter space that fails to satisfy the condition $M/T_{c}<1.6$ for which the high temperature expansion breaks down (see appendix A for details).  Values of $v_{c}(T_c)/T_c>1$ are represented by dark color, with green being $< 1$.

Although there is a reasonable region of the parameter space for which the electroweak baryogenesis can be realized, some of the parameters may spoil the electroweak precision data.
This situation is displayed in Fig. \ref{STPLANE} which is based on a sample of $3000$ models of randomly distributed masses in the range $100<m_{h_0}<150$, $250<m_{A}<300$, $120<m_{H}<170$, $190<m_{H_\pm}<240$, along with one family of mirror particles ($n_g=1$) with  $50<m_{N}<100$,  $115<m_{E}<165$, $85<m_{U}<135$ and $50<m_{D}<100$. The figure shows that many models can lead to electroweak baryogenesis and yet respect the electroweak precision data. However, as we discussed in section 2, using three generations of mirror particles ($n_g=3$) puts strong constraints on the masses of the Higgs as well as  mirror particles. Although  moderate values of the latter may not have dramatic effects, we find that the large Higgs masses needed to adjust the $T$ parameter spoils baryogenesis.

\section{Dark matter}

Mirror particles provide a very interesting possibility as dark matter candidates. Indeed the neutral lightest mirror particle (LMP) is one of these choices. In the following we take the mirror neutrino $N$ to be the LMP. If the LMP decays too fast into normal quarks and leptons, then there is no connection of the mirror particles with dark matter. However, if the LMP is stable or long lived, then it can play the role of dark matter, especially in the context of Inelastic Dark Matter \cite{TuckerSmith:2001hy,Cui:2009xq, Chang:2008gd}.

In order to understand the lifetime constraints, consider the decay $N \longrightarrow \mu^-\,e^+\,\nu_e$ which results due to the mixing terms in (\ref{mixing Lagrangian}). This process can be expressed in terms of the mixing angles $\theta_{1,2\, m}$
\begin{eqnarray}
\nonumber
N&=&N'\cos(\theta_{1\,m})+\nu'\sin(\theta_{1\,m})\,,\\
E&=&E'\cos(\theta_{2\,m})+e'\sin(\theta_{2\,m})\,.
\end{eqnarray}
Neglecting $m_\nu$, $m_e$, and $m_\mu$ compared to $m_N$, and assuming  $\theta_{1\,m}\approx\theta_{2\,m}<<1$ we obtain the life time
\begin{equation}
\tau_N=\frac{12 (8\pi)^3}{m_N\theta_m^2}\left(\frac{m_N g_2}{m_W}\right)^4\,.
\end{equation}
This life time can be used to set bounds on  $\theta_m$. A first bound comes from the LEP data. If the masses of the $N$ particles are between $50$~GeV and $100$~GeV, they would have been discovered at LEPII if they decayed through the weak interactions to normal matter. However, if they were effectively stable, this bound does not apply. If we consider heavy neutrino pair production $e^+\,e^-\longrightarrow N\,\bar N$ at a CM energy $200$ GeV we find that the $N$ will live long enough to exit the detector if $\theta_m \lesssim 10^{-7}$. The other bound comes from considering $N$ a dark matter candidate. Assuming these particles were created early in the Universe, and they do not decay until now we find $\theta_{m} \approx 10^{-32}$.

The relic abundances of the mirrors is not readily predicted. Since as we saw in the last section, we need a form of leptogenesis or baryogenesis in order to understand the baryon asymmetry, the same mechanism would be expected to produce a net asymmetry in the mirror sector. A leptogenesis scenario could lead to a non-zero net value of the mirror lepton quantum number. However, the magnitude of that asymmetry is unknown because it relies on CP violating and lepton number violating parameters of the underlying theory.

The mirror particles are weakly interacting, and hence are candidates for WIMPs. However, the standard weak cross section of a heavy neutral particle initially appears too large. The effective scattering cross section of an SU(2) neutral mirror fermion $\psi_{m}$, through $Z^0$ exchange with a nucleon, is given by \cite{Jungman:1995df}
\begin{equation}
\sigma_{p,n}=\frac{\mu_{p,n}^2}{\pi}\left(\frac{g_2}{M_Z}\right)^4\left(g_v^{\psi_m}g_v^{p,n}\right)\,
\end{equation}
where $\mu_{p,n}$ is the reduced mass of the $\psi^m$-nucleon and $g_v^p=2g_v^u+g_v^d$, and $g_v^n=g_v^u+2g^d_v$. The couplings of the various components are $g_v^{LMP}=1/2$, $g_v^p=1/4-\sin^2\theta_W$, and $g_v^n=-1/4$. Hence the cross sections is
\begin{equation}
\sigma=\frac{G_F^2}{2\pi}\mu_n^2\approx 7.44\times 10^{-39} cm^2\,
\label{weakcrosssection}
\end{equation}
where $G_F$ is the Fermi constant, and $\mu_n\approx 0.939$ GeV is the neutron-Dark matter reduced mass assuming that the latter is much larger than the former.

This standard weak cross section is large and has been excluded by many of the dark matter direct-search  experiments like CDMS and XENON. However, the situation can be rescued by the observation that a Majorana mass for the heavy neutral $N$ splits the mass eigenstates and decreases the elastic cross section - this observation is the basis for the theory of Inelastic Dark Matter\cite{TuckerSmith:2001hy,Cui:2009xq, Chang:2008gd}. The Dirac fermion is split into a pair of Majorana states by a small Majorana mass. Because the weak scattering off of nuclei involves a tansition from one Majorana state to another, as the Majorana mass term increases there exists an increasing energy threshold for the scattering to occur. Dark matter in the galactic halo nay not have enough energy to overcome this threshold.

At this stage, the appropriate parameters depends on whether we accept the results of the DAMA/NaI and DAMA/LIBRA experiments as a signal of dark matter or not. These experiments observe an annual modulation signal in their detectors and the validity of this signal as a sign of dark matter is still controversial. If we do not accept the DAMA results, then we can reduce the mirror dark matter cross section to acceptable values simply by choosing the mirror Majorana mass ${\cal M}_L^N > 150$ keV.

However the situation is more complicated if we do consider the DAMA results as valid signals of dark matter. The Inelastic Dark Matter picture has the possibility of explaining the annual modification if the Majorana mass is chosen such that the threshold effects vary over the time of the year due to the Earth's motion through the dark matter cloud. This effect depends also on the relic density and the mass of the Dark Matter candidate, as well as the basic cross section. Recent analyses \cite{TuckerSmith:2001hy,Cui:2009xq, Chang:2008gd} show that in general higher masses and lower intrinsic cross sections are generally preferred. However, there is a window where the LMP has a mass around $70$~GeV where the weak cross-section of Eq. (\ref{weakcrosssection}) is allowed, so that the mirror model is also marginally able to explain this result also.

In summary, there are portions of vectorial SM parameter space which are plausible for the use of the lightest mirror particle as the Dark Matter candidate.

\section{LHC phenomenology}

There exists many phenomenological studies that are relevant for the vectorial version of the Standard Model. The key feature of the model is that it contains 3 generations of mirror particles. The constraints of the precision electroweak parameters force the masses for these particles to be in a rather small corner of parameter space. Most importantly, this corner contains a light $N$ particle - the neutral lepton which acts as a right handed partner of the neutrinos - and also a light $D$ quark - this particle being the mirror of the down quark.

There is a basic dichotomy in the search strategies depending on whether the mirrors mix with the regular quarks and leptons or not. If the mirrors are stable, or so long lived that they appear stable in accelerator-based experiments, the searches will be same as those for any stable new particle. On the other hand, if the mixing with the normal particles is such that the mirrors decay in the detector, they will be searched for by studying their decay products.
\begin{figure*}[ht]
\leftline{
\includegraphics[width=1\textwidth]{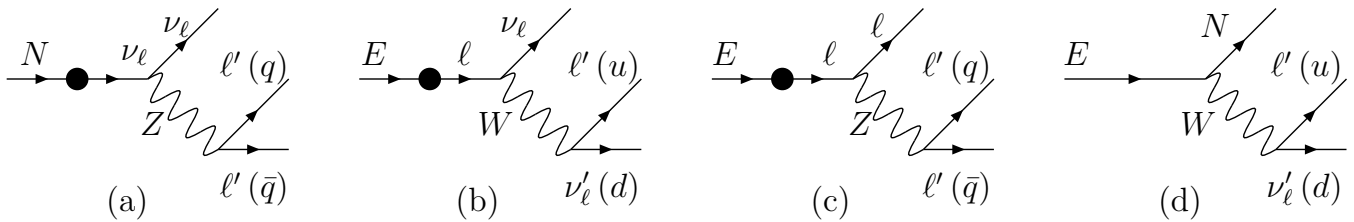}
}
\caption{Decay channels of the leptonic mirror particles which proceed through mixing with the regular leptons. The circle represents a mass mixing insertion from (\ref{mixing Lagrangian}) .}
\label{decaychannels}
\end{figure*}

Search strategies for new stable particles have been extensively studied for the LHC \cite{stable}. Basically, the conclusion is that the LHC will readily be able to see mirror particles with these masses and couplings. Already the Tevatron is getting close to probing this parameter space \cite{limit}. The lightest mirror quark is the $D$ and it will be pair produced with a strong interaction cross section. It will primarily form bound states $D\bar{u}$ and $D\bar{d}$. Heavy quark symmetry suggest that the relative masses would be similar to that of the $B_u,~B_d$ system such that the neutral state would be slightly heavier (due primarily to the $d-u$ mass difference) with both states being stable. The charged state makes a particularly good signal, producing a visible track which exits the detector. The lightest mirror
lepton is the neutral $N$. It is produced in pairs through the $Z\to \bar{N}+N$ coupling. The charged current coupling through $W\to \bar{N}+E$, with the subsequent decay $E\to N+W$ will also populate the $\bar{N}N$ final state but a higher energy threshold. The charged lepton can be pair produced electromagnetically, $E^+E^-$, with subsequent decay down to the neutral states. All of these cases would be seen in the missing energy searches.

The phenomenology for new mirror particles that mix with the normal particles is similar to that of a 4th generation. The fact that the mirrors have right handed weak couplings is less important than the uncertainties in the masses of the new particles. A recent overview is provided by \cite{Kribs:2007nz} and many further references can be found in \cite{unstable}. The leptonic decay channels will be relatively clean - the important diagrams are shown in Fig. \ref{decaychannels}. The mass insertions from Eq (\ref{mixing Lagrangian}) provide the equivalent of CKM mixing factors - however, there is also mixing in the neutral current sector as is evidenced by the $N$ decay diagram. For $N$ decay, the $Z^0$ will be off-shell for the lower end of the favored $N$ mass range, but it will produce an on-shell $Z^0$ at higher masses. The di-lepton plus missing energy signal will be particularly striking.  For $E$ decay, the $Z$ or $W$ will always be on-shell. For mirror quark decays, the signals will be the same as the decay of a heavy fourth generation $b'$ quark.




\section{Summary}

We have explored the remaining parameter space of the minimal vectorial version of the Standard Model. This construction initially has only vectorial gauge currents, but the Higgs sector introduces differences in the chiral structure. The model has to be fine-tuned in order to satisfy the precision electroweak constraints. However, there is still available parameter space consistent with experiment. The constraints from electroweak physics constrain the model and make it easier to understand the remaining physics, which otherwise would be clouded by a wide range of parameters. In particular, we saw that the model is not able to explain weak-scale baryogenesis, but can be a potential dark matter candidate if the Majorana mass terms are appropriate.

The model will be well tested at the LHC. Our analysis indicates that the model must have a light neutral lepton $N$ and a light mirror quark $D$. These results are shown in Sec 3. There are also additional physical Higgs bosons. The masses of all these particles should be readily probed by the LHC. Indeed, the fermion states in particular should be found in the early operation of the LHC. This model may be the first one confirmed or ruled out by the LHC.

\section*{Acknowledgement}
This work has been supported in part by the NSF grant PHY- 055304 and in part by the Foundational Questions Institute. We would like to thank  Paul Langacker and Jens Erler for providing the data file used to plot the 90 $\%$ C.L. contours of the S,T,U parameters. We also thank Gene Golowich, Barry Holstein and Daniel Wyler for conversations.

%
\appendix
\renewcommand{\theequation}{A\arabic{equation}}
  \setcounter{equation}{0}  
  \section*{Appendix A: Discrete symmetries of the model}

In table (\ref{symmetries}) we display the different discrete groups that are respected by the Higgs and Yukawa Lagrangian, along with the bare mass terms allowed by each group (see the section about mixing terms). It is not difficult to see that the set of the discrete groups, S1 to S14, form a group ${\cal G}$. The groups S1 to S6 prevent the mixing of $\Phi_1$ and $\Phi_2$, but allow for some bare mass terms. In contrast, S7 to S10 kill all bare mass terms, however they do not prevent the mixing. Finally, the groups S11 to S14 are trivial in the sense that they do not prevent the mixing  nor they eliminate any of the possible bare masses. Notice also that the set $\{ \mbox{ S11, S12, S13, S14} \}$ is a proper subgroup of ${\cal G}$.

\begin{table*}
\centerline{
\begin{tabular}{|c||c|c|c|c|c|c|c|c|c|c|c|c|c|c|}
	\hline
          & S1 & S2 &S3 & S4 & S5 & S6 & S7 & S8 & S9 & S10 & S11 & S12 & S13 & S14 \\ \hline \hline
 $\Phi_1$ & +  & +  & +  & -  & -  & -  & +  & +  & -  & -  & - & - & +  & +  \\ \hline
 $\Phi_2$ & -  & -  & -  & +  & +  & +  & +  & +  & -  & -  & - & - & +  & +   \\ \hline
 $\psi_R$ & +  & -  & +  & -  & +  & +  & +  & -  & -  & +  & + & - & -  & +   \\  \hline
 $\psi_L$ & -  & +  & +  & +  & -  & +  & -  & +  & +  & -  & + & - & -  & +   \\ \hline
 $c_R$    & -  & +  & +  & -  & +  & -  & -  & +  & -  & +  & - & + & -  & +  \\ \hline
 $c_L$    & -  & +  & -  & -  & +  & +  & +  & -  & +  & -  & - & + & -  & +  \\ \hline
 $f_R$    & -  & +  & +  & -  & +  & -  & -  & +  & -  & +  & - & + & -  & +  \\ \hline
 $f_L$    & -  & +  & -  & -  & +  & +  & +  & -  & +  & -  & - & + & -  & +  \\ \hline
 \hline
 bare m & $c,f$ & $c,f$ & $\psi$ & $c,f$ & $c,f$ & $\psi$ & non & non & non & non & c,f,$\psi$ & c,f,$\psi$ & c,f,$\psi$ & c,f,$\psi$  \\
	\hline		
\end{tabular}
}
\caption{The different discrete groups under which (\ref{Vdouble}) and (\ref{Yukawa Lagrangian}) are invariant. The last row displays the bare mass terms  allowed by each group.}
\label{symmetries}
\end{table*}


 \renewcommand{\theequation}{B\arabic{equation}}
  \setcounter{equation}{0}  
  \section*{Appendix B: Phase transition calculations}  

In this appendix we work out the details of the phase transition in the presence of two Higgs doublet as well as mirror particles \cite{Cline:2006ts, Cline:1995dg}.

\subsection*{Higgs Spectrum}

To study the Higgs mass spectrum, we first write $\Phi_1$ and $\Phi_2$ as
\begin{eqnarray}
\nonumber
\Phi_{1}&=&\frac{1}{\sqrt{2}}\left( \begin{array}{c}
\psi_1+i\psi_2 \\
 \psi_3+i\psi_4 \\
 \end{array} \right)
 =\frac{1}{\sqrt{2}}\left( \begin{array}{c}
\phi_{1}^{+} \\
 \frac{v}{\sqrt{2}}+\eta_{1}+i\chi_{1} \\
 \end{array} \right)\,\mbox{,}\\
 \nonumber
 \Phi_{2}&=&\frac{1}{\sqrt{2}}\left( \begin{array}{c}
\psi_5+i\psi_6 \\
 \psi_7+i\psi_8 \\
 \end{array} \right)
 =\frac{1}{\sqrt{2}}\left( \begin{array}{c}
\phi_{2}^{+} \\
 \frac{v}{\sqrt{2}}+\eta_{2}+i\chi_{2} \\
 \end{array} \right)\,.
  \label{phi}\\
\end{eqnarray}
Next, we define the charged and neutral Goldstone bosons $G^{\pm}$ and $G^0$, and the charged and neutral fields $H^{\pm}$ and $A_0$
\begin{eqnarray}
\nonumber
\left(\begin{array}{c} \phi_1^{\pm}\\ \phi_2^{\pm}  \end{array}\right)&=&\sqrt{2}\left(\begin{array}{cc} \cos\beta & -\sin\beta\\ \sin\beta &\cos\beta \end{array}  \right)\left(\begin{array}{c}G^\pm\\H^\pm  \end{array}\right)\,, \\
\label{GH}
\left(\begin{array}{c} \chi_1\\ \chi_2^{\pm}  \end{array}\right)&=&\left(\begin{array}{cc} \cos\beta & -\sin\beta\\ \sin\beta &\cos\beta \end{array}  \right)\left(\begin{array}{c}G_0 \\ A_0  \end{array}\right)\,,
\end{eqnarray}
where $\phi_{1,2}^{-}=\left(\phi_{1,2}^{+}\right)^{\dagger}$.
The mass matrix is given by $M_{ij}=\partial_i\partial_jV(\Phi_1,\Phi_2)|_{\Phi_{1,2}=\langle\Phi_{1,2}\rangle }$. Substituting (\ref{GH}) into (\ref{Vdouble}) we obtain
\begin{eqnarray}
\nonumber
m_{A}^2&=&-\mu_1^2+\frac{1}{4}\left(h_1+h_2-6h_3+\lambda_1 \right)v^2\,, \\
\label{mA0andMhpm}
m_{H_\pm}^2&=&-\mu_1^2+\frac{1}{4}\left(-h_1+h_2-2h_3+\lambda_1 \right)v^2
\end{eqnarray}
and
\begin{equation}
M_{\eta_1,\eta_2}^2=\left( \begin{array}{cc}
M_{11}^2 & M_{12}^2 \\
 M_{12}^2 &M_{22}^2 \\
 \end{array} \right)\,
\end{equation}
where
\begin{eqnarray}
\nonumber
M_{11}^2&=&M_{22}^2=-\mu^2+\frac{1}{4}(3\lambda_1+h_1+h_2+2h_3)v^2\,,\\
\label{m11m12}
M_{12}^2&=&\frac{1}{2}\left(h_1+h_2+2h_3 \right)v^2\,,
\end{eqnarray}
The eigenvalues of $M_{\eta_1,\eta_2}$ determine the mass of the CP-even fields $(h_0,H_0)$
\begin{eqnarray}
\nonumber
m_{h}^2&=&-\mu_1^2+\frac{3}{4}\left(h_1+h_2+2h_3+\lambda_1\right)v^2\,,\\
\label{mhoandmH0}
m_{H}^2&=&-\mu_1^2-\frac{1}{4}\left(h_1+h_2+2h_3-3\lambda_1\right)v^2\,,
\end{eqnarray}
with rotation angle $\alpha=\pi/4$ in the $\eta_1-\eta_2$ plane.

\subsection*{Loop corrections and ring diagrams}

 By construction, we take $m_{h}$ to be the lightest Higgs boson, while other heavy Higgs bosons as well as fermions run in loops of $h$. To this end we take $\Phi_1=\Phi_2=(0,\phi)^T/2$ in (\ref{Vdouble}) to find
\begin{equation}
V(\phi)=\lambda(\phi^2-v^2)/4\,.
\end{equation}
At one loop and at zero temperature the Higgs-potential, involving all the field-dependent particle masses, reads
\begin{eqnarray}
\nonumber
V_{\mbox{\scriptsize eff}}(\phi)&=&\frac{\lambda}{4}\left(\phi^2-v^2 \right)^2+\frac{1}{2}A\phi^2\\
\nonumber
&+&\frac{1}{64\pi^2}\sum_{B} n_{B}M^4_{B}(\phi)\left(\log\frac{M^2_{B}(\phi)}{\mu^2}-\frac{3}{2}\right)\\
\nonumber
&-&\frac{1}{64\pi^2}\sum_{F} n_{F}M^4_{F}(\phi)\left(\log\frac{M^2_{F}(\phi)}{\mu^2}-\frac{3}{2}\right)\,,\\
\label{effective potential}
\end{eqnarray}
where $n_{B}=1$ and $n_{F}=4$ are the number of degrees of freedom for scalars and fermions, respectively. The constants $A$ and $\mu$ can be fixed by requiring that the one loop correction does not alter the value of the VEV, i.e. $V'_{\mbox{\scriptsize eff}}(v)=0$ and $V''_{\mbox{\scriptsize eff}}(v)=m_{h_0}^2$.

At finite temperature, there is additional one-loop contribution that is given for bosons and fermions, at sufficiently high  temperatures, i.e. at $M/T<1.6$, by
\begin{eqnarray}
\nonumber
V_{b,T}=&-&\frac{\pi^2 T^4}{90}+\frac{M^2T^2}{24}-\frac{M^3T}{12\pi}\\
\nonumber
&-&\frac{M^4}{64\pi^2}\left(\log\frac{M^2}{T^2}-c_b\right)\\
\nonumber
V_{f,T}=&-&n_{F}\left[-\frac{7\pi^2T^4}{720}+\frac{M^2T^2}{48}\right.\\
\label{boson and fermion potential}
&+&\left.\frac{M^4}{64\pi^2}\left(\log\frac{M^2}{T^2}-c_f \right)\right]\,,
\end{eqnarray}
where $c_b=5.40$ and $c_f=2.63$. Moreover, one can correct the Higgs one-loop potential by adding to the loop all the ring diagrams. This can be achieved by replacing $M^2\rightarrow M^2(\phi,T)$ in (\ref{boson and fermion potential}). In general, $M^2(\phi,T)$ takes the form $M^2(\phi,T)=M^2(\phi)+\alpha T^2$, for some coefficient $\alpha$ that depends on the masses of  $W$, $Z$ and fermions. Setting this coefficient to zero compensates for the missing active parameter space which is a result of neglecting the low temperature  expansion in our analysis.

The field dependent mass can be found by redoing the steps that lead to (\ref{mA0andMhpm}) and (\ref{mhoandmH0}) after replacing $v\rightarrow \phi$, and expressing $\mu_1$, $\lambda_1$, $h_{1,2,3}$ in terms of the Higgs masses computed at the VEV
\begin{equation}\label{corrected Higgs masses}
m_{i}^2(\phi)= \frac{1}{2}\left(-m_{h_0}^2+(m_{h_0}^2+2m_{i}^2)\frac{\phi^2}{v^2} \right)\,,
\end{equation}
where $ i=h,G_0,G_\pm,A,H,H_\pm$. Finally, the fermion and gauge bosons corrected masses read
\begin{eqnarray}
\nonumber
m_{F}(\phi)&=&m_{F}\phi/v\,, \quad m_{W}^2(\phi)=g_2^2\phi^2/4\,,\\
\label{corrected fermion masses}
m_{Z}^2&=&(g_1^2+g_2^2)\phi^2/4\,.
\end{eqnarray}
%

\subsection*{Phase transition}

Substituting (\ref{corrected Higgs masses}) and (\ref{corrected fermion masses}) into (\ref{effective potential}) and (\ref{boson and fermion potential}),  we obtain
\begin{equation}\label{compact vtotal}
V_{\mbox{\scriptsize total}}(\phi,T)=\frac{1}{4}\Gamma_{4}(T)\phi^4+\frac{1}{3}\Gamma_{3}(T)\phi^3+\frac{1}{2}\Gamma_{2}(T)\phi^2\,,
\end{equation}
where
\begin{eqnarray}
\nonumber
\Gamma_2(T)&=&-\lambda v^2+A-\frac{c_b-1.5}{64\pi^2}m_{h_0}^2\omega_1\\
\nonumber
&+&\left[\frac{\sum_q m_{q}^2}{2v^2}
+\frac{\sum_l m_{l}^2}{6v^2}+\frac{\omega_1}{24}+ \left(g_1^2+3g_2^2 \right)/2  \right]T^2\\
\nonumber
&-&\frac{\omega_1}{64 \pi^2}m_{h_0}^2\log\frac{T^2}{\mu^2}\,,\\
\nonumber
\Gamma_4(T)&=&\lambda+\frac{\omega_3}{64\pi^2}\left(\log\frac{T^2}{\mu^2}-1.5+c_b \right)\\
\nonumber
&-&\left(\frac{3\sum_{q}m_{q}^4+\sum_{l}m_{l}^4}{4\pi^2 v^4} \right)\left(\log\frac{T^2}{\mu^2}-1.5+c_f \right)\\
\nonumber
\Gamma_{3}(T)&=&-\frac{T}{8\sqrt{2}\pi}\omega_2\,,\,\,\,\omega_1=\sum_{i}\left(m_{h_0}^2+2m_{i}^2 \right)/v^2\,,\\
\nonumber
\omega_2&=&\sum_{i}\left(m_{h_0}^2+2m_{i}^2 \right)^{3/2}/v^3\\
\nonumber
&+&3\sqrt{2}\left(2g_2^3+(g_1^2+g_2^2)^{3/2}\right)/4\,,\\
\nonumber
\omega_3&=&\sum_{i}\left(m_{h_0}^2+2m_{i}^2 \right)^{2}/v^4\\
\label{Gammacoeff}
&+&3\left(2g_2^4+(g_1^2+g_2^2)^2 \right)/4\,,
\end{eqnarray}
where the sum is over $h,G_0,G_\pm,A,H,H_\pm$,  and we have used the approximation $m_{i}^2(\phi,T)\approx  \left(m_{h_0}^2+2m_{i}^2\right)\phi^2/2v^2$ in $\Gamma_3$.
The first order phase transition happens when the non-trivial minimum $v_c(T_c)$ in the potential becomes degenerate with the minimum at the origin. Hence, (\ref{compact vtotal}) can be written as
\begin{equation}\label{final compact vtotal}
V_{\mbox{\scriptsize total}}(\phi,T)=\frac{1}{4}\Gamma_4(T) \phi^2(\phi-v_c(T))^2\,.
\end{equation}
Comparing the coefficients in (\ref{compact vtotal}) and (\ref{final compact vtotal}) we obtain
\begin{eqnarray}
v_c(T_c)&=&-\frac{2\Gamma_{3}(T_c)}{3\Gamma_4(T_c)}\\
\Gamma_2(T_c)&=&\frac{\Gamma_4(T)v_c(T_c)^2}{2}\,,
\end{eqnarray}
from which we solve for $T_c$ and $v_{c}(T_c)$.

\end{document}